\documentclass[aps,prb,showpacs,twocolumn,groupedaddress,floatfix]{revtex4}%
\usepackage{subfigure}
\usepackage{amsmath}
\usepackage{graphicx}
\usepackage{rotating}
\usepackage{amsfonts}
\usepackage{amssymb}%
\providecommand{\U}[1]{\protect\rule{.1in}{.1in}}
\begin{document}
\title{Superconductivity and Magnetism in BiOCuS}
\author{I.I. Mazin}
\affiliation{Code 6393, Naval Research Laboratory, Washington, D.C. 20375}
\date{Printed on \today}

\begin{abstract}
BiOCu$_{1-y}$S was recently reported to superconduct at $T\approx5.8$ K at
$y\approx0.15.$ Band structure calculations indicate that the stoichiometric
BiOCuS is a band insulator. In this paper, I show that the hole-doped (whether
in the virtual crystal approximation or with actual Cu vacancies) BiOCuS is on
the verge of a ferromagnetic instability (\textit{cf.} Pd metal), and
therefore a conventional superconductivity with $T_{c}\sim6$ K is quite
unlikely. Presumably, the hole-doped BiOCuS is another example of
superconductivity mediated by spin-fluctuations.

\end{abstract}

\maketitle

Recently, Ubaldini $et$ $al$ \cite{Dirk} reported that in the quaternary
compound BiOCu$_{1-y}$S superconductivity at $T_{c} = 6$ K can be induced by a
small Cu deficiency ($y\sim15\%),$ although not by an electron doping (when
substituting O with F). Although another similar study has not confirmed the
superconductivity \cite{Ind}, this result has triggered some interest in the
community\cite{Ind,SI}. I will show below that this interest is well
justified, for this material is very unlikely to be a conventional s-wave
superconductor, and, if superconducting, is probably another example of a
spin-fluctuation mediated superconductivity.

BiOCuS crystallizes in the well known tetragonal ZrCuSiAs
structure\cite{Ind,str}, recently made famous by some Fe-based
high-temperature superconductors (although there is no commonality whatsoever
between the latter and the material in question). The lattice parameters
reported in Ref. \onlinecite{Dirk} are 3.8726 and 8.5878 \AA , in Ref. \onlinecite{str}
3.8691 and 8.5602 \AA , and in Ref. \onlinecite{Ind} 3.868 and 8.557 \AA . The
internal coordinates of Bi and S, respectively, are reported as 0.14829 and
0.6710 (Ref. \onlinecite{str}) and 0.151 and 0.648 (Ref. \onlinecite{Ind}).

In my calculations I used the structure reported in Ref. \onlinecite{str}. The
variation of crystallographic parameters withing these limits does not affect
any conclusions of this paper.

\begin{figure}[ptb]
\includegraphics[width=0.95 \linewidth]{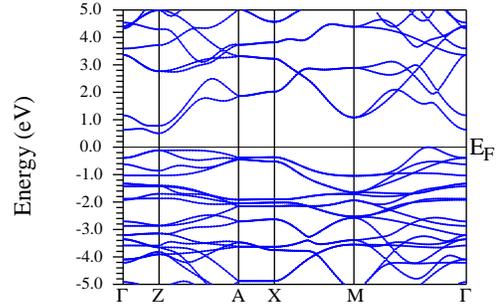}\caption{Band structure of
BiOCuS near the Fermi level. Note a flat band in the $Z-A$ direction.}%
\label{bands}%
\end{figure}

\begin{figure}[ptb]
\includegraphics[width=0.95 \linewidth]{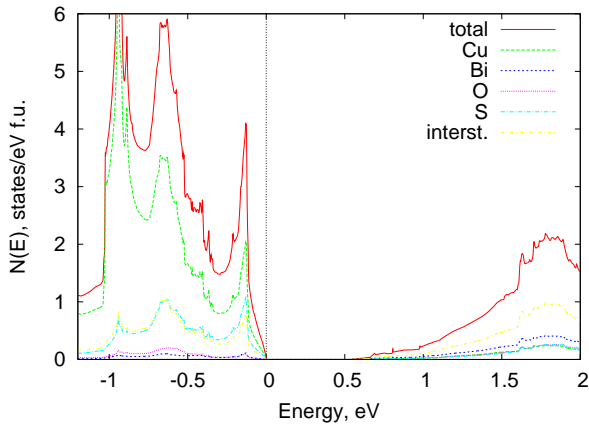}\caption{Density of states of
BiOCuS near the Fermi level. Note a peak right below the gap, and the leading
contribution of the Cu states below, and the interstitial states above the
band gap. (color online)}%
\label{DOS}%
\end{figure}

\begin{figure}[ptb]
\includegraphics[width=0.49 \linewidth]{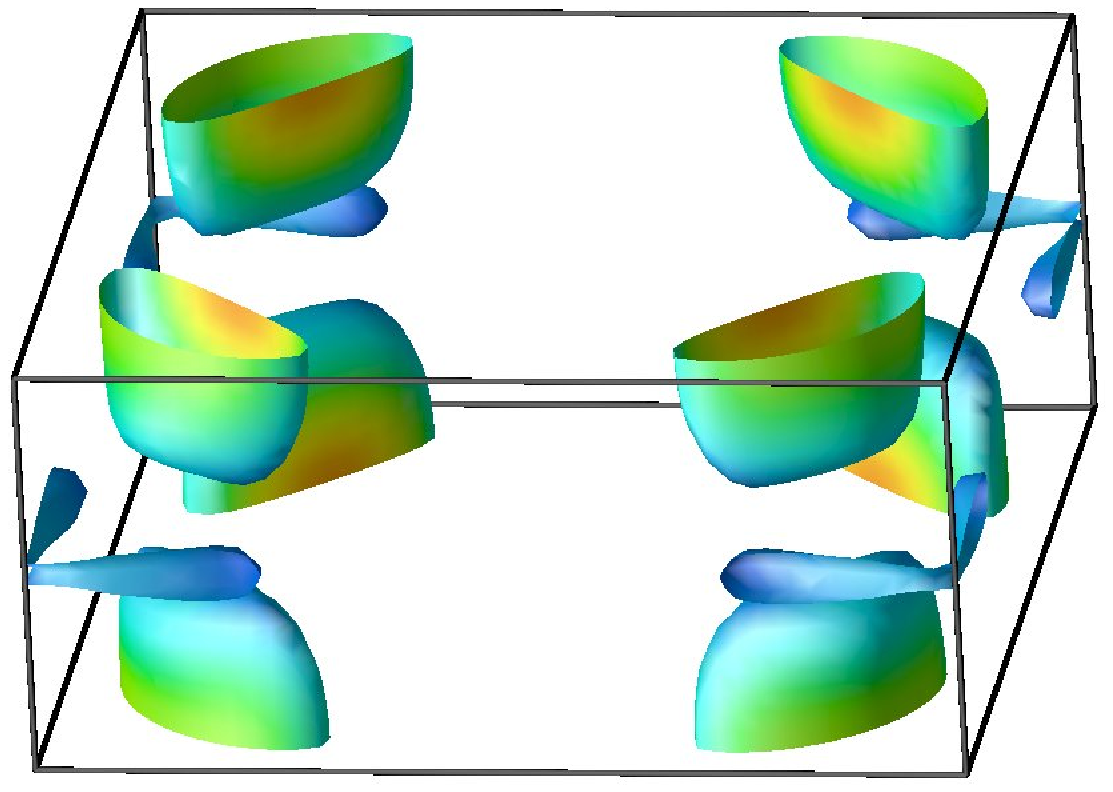}
\includegraphics[width=0.49 \linewidth]{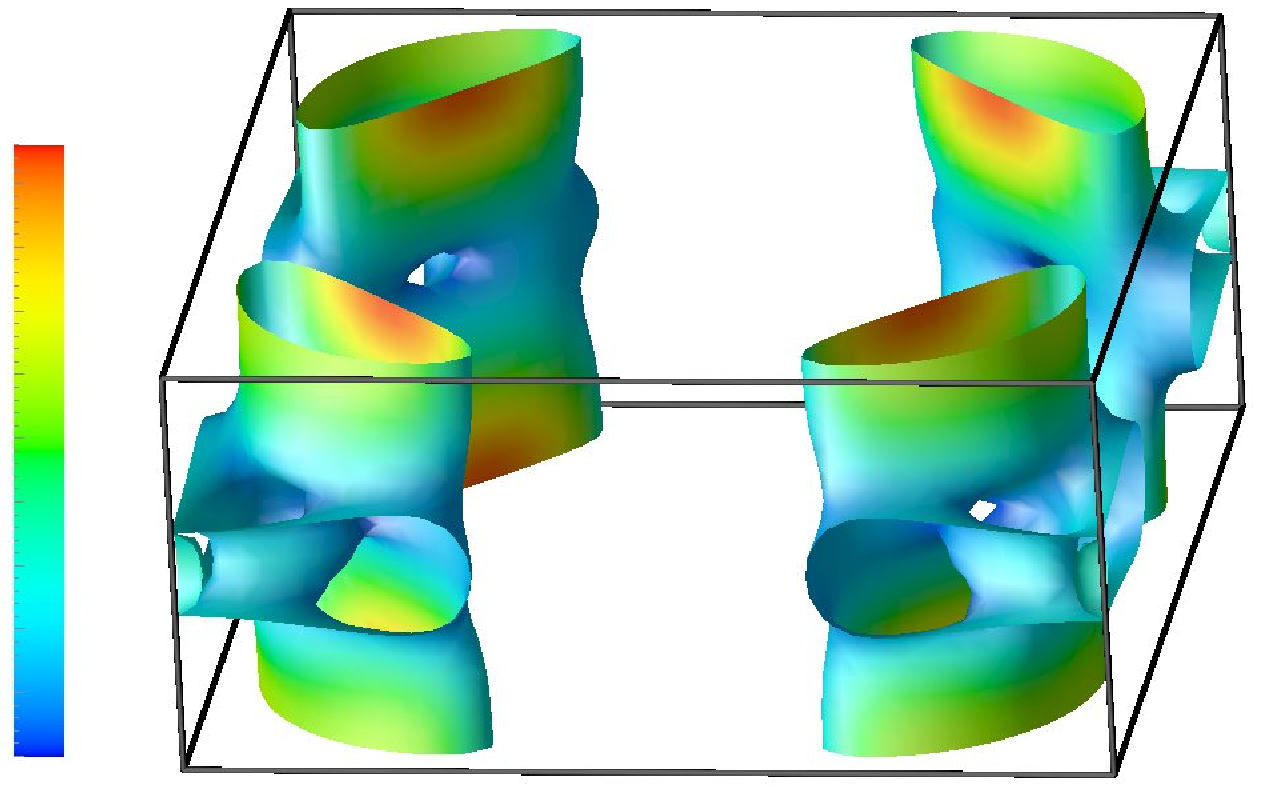}\caption{ (color online)
Left panel: The Fermi surface of BiOCu$_{1-y}$S at $E_{F}$  120 meV below the
gap, where a sharp peak in DOS starts. Note a cross-shaped pocket emerging
along the $Z-A$ direction, responsible for this peak. For comparison, in the
right panel the Fermi surface at $E_{F}=150$ meV ($y \sim 0.15$) is shown, where the peak is
already fully developed. Both Fermi surfaces are colored according the the
absolute value of the Fermi velocity using the same scale (blue is zero).}%
\label{FS}%
\end{figure}

In agreement with the previous first principles calculations\cite{SI,str} I
found\cite{method} that in the scalar-relativistic approximation the stoichiometric
compound is a band insulator, with a direct gap underestimated compared to the
measured optical gap\cite{str}, as common in the density functional
calculations. Adding spin-orbit interaction on Bi reduces the gap even further, 
from 0.75 to 0.55 eV, by affecting the unoccupied states. The states right
below the gap are formed predominantly by the Cu $xz$ and $yz$ orbitals, while
the empty states above the gap have quasi-free-electron character (the
interstitial region, being $\approx$60\% in volume, contributes up to 50\% of
the density of states of these bands). In agreement with, especially, the full
potential all-electron calculations of Ref. \onlinecite{SI}, I found a flat band
right at the top of the valence band at the Z point (Fig. \ref{bands}),
essentially not dispersing up to half-way between Z and X\cite{note2}. This
band results in a small but very sharp density of states (DOS) peak right
below the band gap (Fig. \ref{DOS}), with the total weight of slightly less
than 0.3 electron per copper, and a maximum DOS of 4.1 states/eV. f.u. at 130
meV below the gap (corresponding to 0.10--0.12 hole/Cu). One can see that this
peak indeed originates from the flat band along $Z-A$, by looking at the Fermi
surface for $E_{F}$ set at 120 meV below the band gap, where the peak just
starts (approximately 0.07 hole/Cu doping). This feature appears as
cross-shaped Fermi surface pockets emerging from Z (Fig. \ref{FS}).



\begin{figure}[ptb]
\includegraphics[width=0.95 \linewidth]{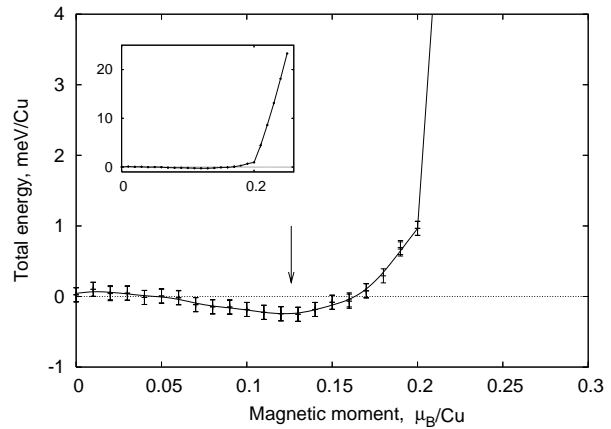}\caption{Fixed spin
moment calculations for BiOCu$_{1-y}$S for $y=0.2$. The self-consistent
solution is indicated by the arrow. Note that the energy is essentially
independent of the magnetic moment up to the half-metallic magnetization of
$y$ $\mu_{B}$/Cu. The inset shows a linear growth of the total energy when
magnetization exceeds the half-metal limit.}%
\label{FSM}%
\end{figure}

Given that Cu d orbitals have sizeable Hund rule coupling this result suggests
that Cu ions under 10-20\% hole doping should be magnetic. Indeed,
ferromagnetic (antiferromagnetic calculation, however, were invariably 
converging to zero magnetization\cite{note3}) calculations for
$y=0.2$ in the virtual crystal approximation (0.2 hole per Cu) converge to a
stable solution with the magnetic moment $M\approx0.12-0.13$ $\mu_{B}$ per Cu,
but with hardly any energy gain. Correspondingly, fixed spin moment calculations (Fig.
\ref{FSM}) show that in the GGA approximation the energy, within the
computational accuracy, does not depend on magnetization up to $M=y$ $\mu_{B}%
$. Note that this magnetization would correspond to a half-metallic state
similar to that in the Co-doped FeS$_{2}$\cite{CoS2} (in fact the physics of the
formation of a magnetic state is quite similar here). This means that the
calculated spin susceptibility is essentially infinite, a situation extremely
close to Pd metal, where GGA calculations also yield infinite
susceptibility\cite{Pd}. As in Pd, or, for that matter, in Fe pnictides,
fluctuations beyond the mean-field level should suppress magnetization and
reduce the susceptibility to a large, but finite value. But, again as in Pd,
strong spin fluctuations are bound to destroy any conventional s-wave
superconductivity, particularly given that with this relatively modest carrier
density electron-phonon coupling cannot be too strong. 

This leads us to the conclusion that the superconductivity in the hole-doped
BiOCu$_{1-y}$S must be unconventional, and probably magnetically mediated,
although it is not clear yet what particular pairing symmetry can be generated
in this material.

One can question the validity of the virtual crystal approximation, given that
holes are introduced not through substitution, but through Cu vacancies. To
verify that, I have performed supercell calculations using four formula
units with one vacancy, that is to say, Bi$_{4}$O$_{4}$Cu$_{3}$S$_{4},$
corresponding to $y=0.25.$ The calculations converged to a ferromagnetic state
with $M=0.2$ $\mu_{B}$ per formula unit (0.8 $\mu_{B}$ per supercell), again
slightly below the half-metallic limit of 0.25 $\mu_{B},$ thus confirming the
validity of the virtual crystal calculations. Interestingly, the magnetization
was quite delocalized, with Cu ions carrying 0.1 -- 0.13 $\mu_{B},$ and the
rest of the magnetic moment distributed among the sulfur ions and in the
interstitial region.

I thank Dirk van der Marel for stimulating discussions related to this work,
and Peter Blaha for a technical consulation. I also acknowledge funding from
the Office of Naval Research.


\begin{thebibliography}{9}                                                                                                %
\bibitem {Dirk}A. Ubaldini, E. Giannini, C. Senatore, D. van der
Marel\textbf{, }arXiv:0911.5305 (unpublished).

\bibitem {Ind}A. Pal, H. Kishan and V.P.S. Awana, arXiv:0912.0991 (unpublished)

\bibitem {SI}I. R. Shein and A. L. Ivanovskii, Solid State Communications,
\textbf{150}, 640 (2010) 

\bibitem {str}H. Hiramatsu, H. Yanagi, T. Kamiya, K. Ueda, M. Hirano and H.
Hosono Chem., Mater., \textbf{20}, 326 (2008)

\bibitem{method} The full-potential LAPW method, as implemented in the
WIEN2k packege, has been used for all calculations. Up to 840 inequivalent k-points
have been utilized to achieved the self-consistency, with the energy
convergency better then 0.05 meV.

\bibitem {note2}In Ref. \onlinecite{SI}, unconventional notations for the high
symmetry points are used; conventional A ($\pi,0,\pi)$ is called R, and
conventional R ($\pi,\pi,\pi)$ is called A. I am using the standard notations.

\bibitem {note3}I have tried only a checkerboard antiferromagnetic
arrangement, since I do not see any {\it a priori} reason for any other
antiferromagnetic ordering (the fact that the crystal structure coincides with
that of pnictide superconductors is not an argument at all that this material
would assume the same exotic magnetic structure). Of course, this does not
mean that there are no spin fluctuations with a finite wave vector; it only
tells us that if such fluctuations occur, they occur at a wave vector,
different from ${\pi,\pi}$ (in the unfolded Brillouin zone notation).

\bibitem {CoS2}I. I. Mazin, Appl. Phys. Lett. \textbf{77}, 3000 (2000).

\bibitem {Pd}P. Larson, I.I. Mazin, D.J. Singh, Phys. Rev. \textbf{B69},
064429 (2004).
\end{thebibliography}
\end{document}